\documentclass[twocolumn,floatfix, showpacs]{revtex4}

\usepackage[final]{epsfig}
\usepackage{amsmath}
\usepackage{parskip}
 
\begin{document}
 
\title{Can one extract energy loss probability distributions from $R_{AA}$?}
 
\author{Thorsten Renk}
\email{trenk@phys.jyu.fi}
%\author{Kari J.~Eskola}
%\email{kari.eskola@phys.jyu.fi}
\affiliation{Department of Physics, P.O. Box 35 FI-40014 University of Jyv\"askyl\"a, Finland}
\affiliation{Helsinki Institute of Physics, P.O. Box 64 FI-00014, University of Helsinki, Finland}
 
%\preprint{HIP-2006-46/TH}

\begin{abstract}
The nuclear suppression of high transverse momentum $P_T$ hadrons is one of the most striking findings in heavy-ion collision experiments. It has long been recognized that the suppression can be theoretically described by folding the primary parton spectrum with an energy loss probability distribution which is suitably averaged over the collision geometry. However, an interesting problem is to what degree the procedure can be inverted, i.e. given a measurement of the suppression factor $R_{AA}$ with arbitrary precision, can the probability distribution of energy loss be extracted in a model-independent way? In this note, we present a conceptual study of the inversion problem for LHC energies and demonstrate that a measurement of $R_{AA}$ alone is insufficient to determine the distribution, other observables such as $\gamma$-hadron correlations must be taken into account.
 
\end{abstract}
 
\maketitle

The nuclear suppression factor $R_{AA}$, defined as the measured spectrum of high $P_T$ hadrons divided by the value seen in p-p collision scaled with the number of binary collisions is one of the best investigated signatures of jet quenching due to medium induced energy loss \cite{Jet1,Jet2,Jet3,Jet4,Jet5,Jet6}. For pions, it has been measured at RHIC out to 20 GeV $P_T$ by the PHENIX collaboration \cite{PHENIX_R_AA}. From a theory perspective, it has been realized that $R_{AA}$ needs to be computed in a probabilistic framework, not only because the energy loss for a given parton may be probabilistic (cf. e.g. \cite{QuenchingWeights}) but also because the initial vertex is not known experimentally, thus the averaging over the evolving geometry leads to a probability distribution of energy loss (cf. e.g. \cite{JetFlow}).

Combining all probabilistic effects, the effect of jet energy loss may be cast quite generally into a distribution $\langle P(\Delta E, E) \rangle_{T_{AA}}$ which describes the probability to lose energy $\Delta E$ given the initial parton energy $E$ and an unknown initial vertex which is averaged over \cite{gamma-h,Correlations}. 

Under the assumption that the probability distribution for $\Delta E \ll E$ does not strongly depend on $E$, one can reduce the computation to evaluating $\langle P(\Delta E) \rangle_{T_{AA}}$. The medium-modified perturbative production of hadrons can then be computed from the (schematical) expression

\begin{equation}
\label{E-folding}
d\sigma_{med}^{AA\rightarrow h+X} \negthickspace \negthickspace \negthickspace = \sum_f d\sigma_{vac}^{AA \rightarrow f +X} \otimes \langle P(\Delta E)\rangle_{T_{AA}} \otimes
D_{f \rightarrow h}^{vac}(z, \mu_F^2)
\end{equation} 

with $d\sigma_{vac}^{AA \rightarrow f +X}$ the perturbative QCD (pQCD) production cross section for a parton $f$  and $D_{f \rightarrow h}^{vac}(z, \mu_F^2)$ the fragmentation function with momentum fraction $z$ at scale $\mu_F^2$ \cite{KKP,AKK} (see appendix \ref{A-1} and \cite{Correlations} for the explicit expression for Eq.~(\ref{E-folding})). From this one can compute the nuclear modification factor $R_{AA}$  as

\begin{equation}
R_{AA}(p_T,y) = \frac{dN^h_{AA}/dp_Tdy }{T_{AA}({\bf b}) d\sigma^{pp}/dp_Tdy}.
\end{equation}

An intriguing question is now if it is possible to invert Eq.~(\ref{E-folding}) with known $R_{AA}$, assuming full knowledge of parton spectrum and fragmentation function, to determine $\langle P(\Delta E)\rangle_{T_{AA}}$ approximately in a model-independent way.

This can be done by discretizing the integral over $\Delta E$ implicit in Eq.~(\ref{E-folding}) which results in

\begin{equation}
\label{E-Matrix}
R_i(P_T^i) = \sum_{j=1}^n K_{ij}(P_T^i, \Delta E^j) P_j(\Delta E^j)
\end{equation}

where $R_{AA}$ is provided at $m$ discrete values of $P_T$ labelled $R_i$ and $\langle P(\Delta E)\rangle_{T_{AA}}$ is probed at $n$ discrete values of $\Delta E$ labelled $P_j$. $K_{ij}$ is then the calculated $R_{AA}$ for all pairs $(p_T^i, \Delta E^j)$ assuming the pQCD parton spectrum and fragmentation function are known (for the explicit expression see appendix \ref{A-1}).

Eq.~(\ref{E-Matrix}) can be solved for the vector $P_j$ by inversion of $K_{ij}$ for $m=n$, i.e. the experimental resolution of $R_{AA}$ determines the available resolution for the energy loss probability density. However, in general this does not guarantee that the result is a probability distribution. Especially in the face of experimental and theoretical errors and finite numerical accuracy the direct matrix inversion may  permit negative $P_j$ which have no probabilistic interpretation.

Thus, a more promising solution which avoids the above problems is to let $m > n$ and find the vector $P$ which minimizes $|| R - K P||^2$ subject to the constraints $0 \le P_i \le 1$ and $\sum_{i=1}^n P_i = 1$. This guarantees that the outcome can be interpreted as a probability distribution and since the system of equations is overdetermined for $m>n$ errors on individual points $R_i$ do not have a critical influence on the outcome any more.

It has been shown in \cite{gamma-h} that $R_{AA}$ for RHIC kinematics is very insensitive to the shape of the underlying probability distribution. Part of the problem at RHIC is that for a steeply falling parton spectrum, even a small energy loss effectively resembles an absorption of the parton. Therefore, there is some reason to suspect that the inversion problem is badly conditioned. In the following, we study the problem at LHC kinematics where the parton spectrum is harder and larger shifts in parton energy can in principle be probed.

Our setup to test the inversion is as follows: We provide a prediction for $R_{AA}$ at LHC kinematics based on \cite{LHC} sampled at 30 or 50 discrete points between 6 GeV and 400 GeV momentum as input vector $R_i$. We solve for $P_j$ at 10 or 23 discrete points between 0 and 300 GeV energy loss. In addition to these discrete values, we always provide zero (no energy loss) and 2.25 TeV (complete parton absorption) as possibilities and compute $K_{ij}$ accordingly.

\begin{figure}[htb]
\epsfig{file=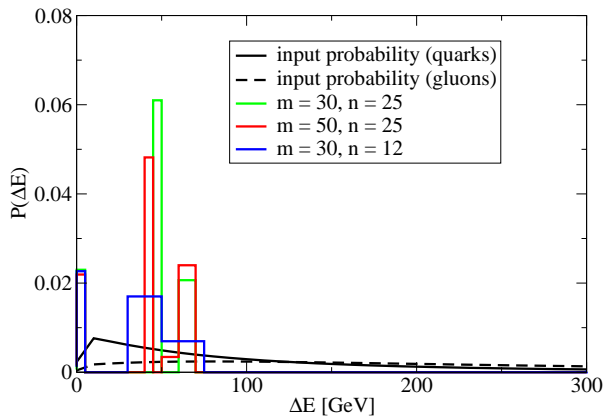, width=7.9cm}
\caption{\label{F-1}Geometry-averaged energy loss probability density for LHC kinematics using a hydrodynamical model of the medium underlying a prediction for $R_{AA}$ \cite{LHC} compared with the probability density extracted from the predicted $R_{AA}$ via numerical
inversion for different numbers $m, n$ (see text).}
\end{figure}

The resulting solution to the vector $P_j$ interpreted as a probability density is shown in Fig.~\ref{F-1} and compared with the (computed) input energy loss probability distribution underlying the prediction of $R_{AA}$ in \cite{LHC}. While details of the distribution depend on the values of $m$ and $n$, the main features of the extracted distribution are rather stable: There is some strength close to zero energy loss, the distribution peaks between 50 and 70 GeV energy loss and the strength in the absorption channel (not seen in Fig.~\ref{F-1}) is about 0.35, i.e. roughly a third of partons is absorbed. However, the most striking observation is that the distribution found from the numerical inversion does not resemble the input distribution closely.

\begin{figure}[htb]
\epsfig{file=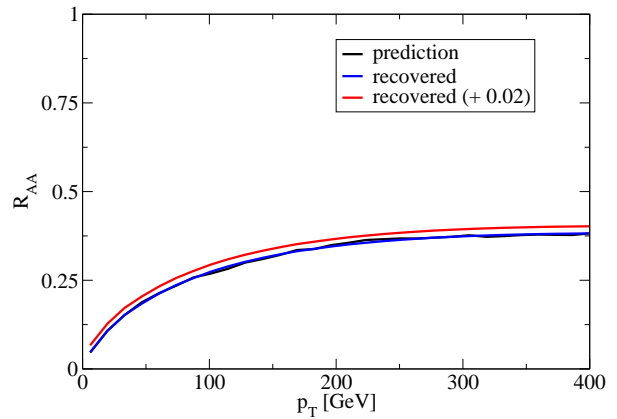, width=7.9cm}
\caption{\label{F-2}$R_{AA}$ as predicted for LHC kinematics using a hydrodynamical model for the medium evolution \cite{LHC} compared with $R_{AA}$ computed from the probability distribution extracted by numerical inversion of $R_{AA}$ (see text). The two curves are virtually identical, hence a shifted recovered curve is provided for comparison.}
\end{figure}

One may ask if the extracted probability distribution is able to describe $R_{AA}$ when inserted back into Eq.~(\ref{E-folding}). This is shown for $m=30, n=25$ in Fig.~\ref{F-2} where we compare the original prediction of $R_{AA}$ with the one recovered from the extracted probability density. As can easily be verified, the results cannot be distinguished, i.e. two very different probability distributions yield an indistinguishable $R_{AA}$.

We are thus faced with a rather asymmetric problem --- the fact that $R_{AA}$ is rather insensitive to the shape of the probability distribution prevents the extraction of anything but gross features of the distribution given only a measurement of $R_{AA}$. Since in reality neither the pQCD spectrum of partons nor the fragmentation function is free of uncertainties and since a measured $R_{AA}$ will have experimental errors, the strategy to invert $R_{AA}$ to find the energy loss probability distribution is simply not feasible in practice.

\begin{figure}[htb]
\epsfig{file=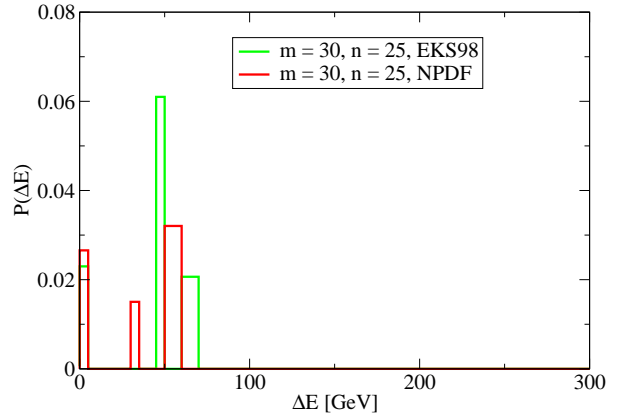, width=7.9cm}
\caption{\label{F-3}Extracted energy loss probability density from the prediction of $R_{AA}$ at LHC under the assumption that the EKS98 nuclear parton distribution or the NPDF nuclear parton distribution is used to compute the folding kernel $K_{ij}$ (the original prediction is based on EKS98).}
\end{figure}

We illustrate the the uncertainty associated with the knowledge of the pQCD kernel $K_{ij}$ in Fig.~\ref{F-3}. Here, we solve Eq.~(\ref{E-Matrix}) using two different sets of nuclear parton distributions to compute $K_{ij}$, i.e. the EKS98 set \cite{EKS98} and the NPDF set \cite{NPDF} (note that the prediction for $R_{AA}$ used as input is based on EKS98). It is clearly seen that the (rather small) uncertainty in the choice of the parton distribution function maps into visible changes in the extracted probability distribution. While gross features and typical energy loss scales are still similar, the differences in detail are substantial.

Thus, while experimental data on $R_{AA}$ are valuable in constraining energy loss models, due to the badly conditioned inversion problem $R_{AA}$ per se does not even measure an energy loss probability distribution, much less a density-distribution of matter (which can only in a model-dependent way be linked to the energy loss distribution) and even less a single parameter characterizing the density. Turning to other, more differential observables (such as $\gamma$-hadron correlation measurements in which the energy of a quark is known before energy loss \cite{gamma-h}) is necessary to achieve the first step towards a model-independent extraction of medium properties from hard probes.

\appendix

\section{Calculating $K_{ij}$}

\label{A-1}

In this appendix, we describe how to compute the inversion kernel $K_{ij}$ as a function of hadron momenta
$P_T^i$ and energy loss $\Delta E^j$ from perturbative QCD (pQCD). 
 
Let $A$ and $B$ stand for the colliding objects (protons or nuclei) and be $y_{1(2)}$ the 
rapidity of parton $k(l)$. The distribution function of a parton type $i$ in $A$ at a momentum 
fraction $x_1$ and a factorization scale $Q \sim p_T$ is $f_{i/A}(x_1, Q^2)$. The distribution 
functions are different for the free protons \cite{CTEQ1,CTEQ2} and protons in nuclei 
\cite{NPDF,EKS98}. The fractional momenta of the colliding partons $i$, $j$ are given by
$ x_{1,2} = \frac{p_T}{\sqrt{s}} \left(\exp[\pm y_1] + \exp[\pm y_2] \right)$.

Inclusive production of a parton flavour $f$ at rapidity 
$y_f$ is found by integrating over either $y_1$ or $y_2$ and summing over appropriate combinations 
of partons,
 
\begin{widetext}
\begin{equation}
\label{E-1Parton}
\begin{split}
\frac{d\sigma^{AB\rightarrow f+X}}{dp_T^2 dy_f}  = \int d y_2 \sum_{\langle ij\rangle, \langle kl  
\rangle} \frac{1}{1+\delta_{kl}} \frac{1}{1+\delta_{ij}} &\Bigg\{ x_1 f_{i/A}(x_1,Q^2) x_2 
f_{j/B}(x_2,Q^2) \bigg[ \frac{d\sigma^{ij\rightarrow kl}}{d\hat{t}}(\hat{s}, \hat{t},\hat{u})  
\delta_{fk} +
\frac{d\sigma^{ij\rightarrow kl}}{d\hat{t}}(\hat{s}, \hat{u},\hat{t}) \delta_{fl} \bigg]\\
+&x_1 f_{j/A}(x_1,Q^2) x_2 f_{i/B}(x_2,Q^2) \bigg[ \frac{d\sigma^{ij\rightarrow kl}}{d\hat{t}}(\hat{s},  
\hat{u},\hat{t}) \delta_{fk} +
\frac{d\sigma^{ij\rightarrow kl}}{d\hat{t}}(\hat{s},\hat{t}, \hat{u}) \delta_{fl} \bigg] \Bigg\} \\
\end{split}
\end{equation}
\end{widetext}
 
where the summation $\langle ij\rangle, \langle kl \rangle$ runs over pairs $gg, gq, g\overline{q}, 
qq, q\overline{q}, \overline{q}\overline{q}$ and $q$ stands for any of the quark flavours $u,d,s$.
Expressions for the pQCD subprocesses $\frac{d\hat{\sigma}^{ij\rightarrow kl}}{d\hat{t}}(\hat{s}, 
\hat{t},\hat{u})$ as a function of the parton Mandelstam variables $\hat{s}, \hat{t}$ and $\hat{u}$ 
can be found e.g. in \cite{pQCD-Xsec}.
 
For the production of a hadron $h$ with mass $M$, transverse momentum $P_T$ at rapidity $y$ and 
transverse mass $m_T = \sqrt{M^2 + P_T^2}$ from the parton $f$, let us introduce the fraction $z$ 
of the parton energy carried by the hadron after fragmentation with $z = E_h/E_f$. Assuming 
collinear fragmentation, the hadronic variables can be written in terms of the partonic ones as
 
\begin{equation}
m_T \cosh y = z p_T \cosh y_f \quad \text{and} \quad m_T \sinh y = P_T \sinh y_f.
\end{equation}
 
Thus, the hadronic momentum spectrum arises from the partonic one by folding with the probability 
distribution $D_{f\rightarrow h}(z, \mu_f^2)$ to fragment with a fraction $z$ at a scale $\mu_f 
\sim P_T$ as
 
\begin{widetext}
\begin{equation}
\label{E-Fragment}
\frac{d\sigma^{AB\rightarrow h+X}}{dP_T^2 dy} = \sum_f \int dp_T^2 dy_f  
\frac{d\sigma^{AB\rightarrow f+X}}{dp_T^2 dy_f} \int_{z_{min}}^1 dz D_{f\rightarrow h}(z, \mu_f^2)  
\delta\left(m_T^2 - M_T^2(p_T, y_f, z)\right) \delta\left(y - Y(p_T, y_f,z)\right)
\end{equation} 
\end{widetext}
 
with
 
\begin{equation}
M_T^2(p_T, y_f, z) = (zp_T)^2 + M^2 \tanh^2 y_f
\end{equation}
 
and 
\begin{equation}
 Y(p_T, y_f, z) = \text{arsinh} \left(\frac{P_T}{m_T} \sinh y_f \right).
\end{equation}
 
The lower cutoff $z_{min} = \frac{2 m_T}{\sqrt{s}} \cosh y$ arises 
from the fact that there is a kinematical limit on the parton momentum; it cannot exceed 
$\sqrt{s}/(2\cosh y_f)$ and thus for given hadron momentum there is a minimal $z$ 
corresponding to fragmentation of a parton with maximal momentum.

The modified spectrum of partons given the energy loss $\Delta E$ computes to

\begin{widetext}
\begin{equation}
\label{E-Eloss}
\frac{d\tilde\sigma_{\Delta E}^{AA\rightarrow f+X}}{dp_T dy_f}   =
\int dq_T d y^*_f d\phi_f^*
\frac{d\sigma^{AB\rightarrow f+X}}{dq_T  dy^*_f}  
\delta(y_f - y^*_f) \delta(p_T - (q_T-\Delta E)) \delta(\phi - \phi_f^*),
\end{equation} 
\end{widetext}

As a side remark, in order to obtain the modified spectrum for a probabilistic energy loss instead of a fixed value of $\Delta E$,  Eq.~(\ref{E-Eloss}) has to be integrated over the probability of energy loss $\int d \Delta E \langle P(\Delta E)\rangle_{T_{AA}}$.
Inserting Eq.~(\ref{E-Eloss}) into Eq.~(\ref{E-Fragment}) instead of $\frac{d\sigma^{AB\rightarrow f+X}}{dp_T^2 dy_f}$ yields
the spectrum of hadrons $\frac{d\tilde\sigma_{\Delta E}^{AA\rightarrow h+X}}{dP_T^2  
dy}(P_T^i,y)$, given the energy loss $\Delta E$.  For given vectors $P_T^i, \Delta E^j$ one can then evaluate

\begin{widetext}
\begin{equation}
\label{E-R_AA}
K_{ij}(P_T^i, \Delta E^j,y) = {\frac{d\tilde\sigma_{\Delta E}^{AA\rightarrow h+X}}{dP_T^2  
dy}(P_T^i,\Delta E^j, y)}/{\frac{d\sigma^{pp\rightarrow h+X}}{dP_T^2 dy}(P_T^i, y)}.
\end{equation}
\end{widetext}

\begin{acknowledgments}
I'd like to thank J\"{o}rg Ruppert and Kari J. Eskola for valuable discussions on the problem. This work was financially supported by the Academy of Finland, Project 115262. 
\end{acknowledgments}

\end{document}